\documentclass[10pt,twoside]{hsqcd}
 \usepackage{epsf,amsmath,amssymb}
  \usepackage[dvips]{graphicx}   
   \graphicspath{{./bakulev.figs/}}   
      
\title{Resummation approach in Fractional APT:\\
        How many loops do we need to calculate?
 \thanks{This work is partially done in collaboration with 
         S.~Mikhailov (JINR) and N.~Stefanis (ITP-II RUB).}
       }

\author{\underline{Alexander~P.~Bakulev} \\ \\
         Bogoliubov Lab. of Theoretical Physics, JINR \\
         Dubna 141980, Russia\\
         E-mail: bakulev@theor.jinr.ru}

\begin{document}
\maketitle

\begin{abstract}
\noindent We give short introduction to the Analytic Perturbation Theory (APT)~\cite{SS} 
 in QCD, describe its problems and suggest as a tool for their resolution 
 the Fractional APT (FAPT)~\cite{BMS-APT,BKS05}. 
 We also describe shortly how to treat heavy-quark thresholds 
 in FAPT and then show how to resum perturbative series in both the one-loop APT and FAPT.
 As applications of this approach we consider the Higgs boson decay $H^0\to b\bar{b}$,
 the Adler function $D(Q^2)$ and the ratio $R(s)$ in the $N_f=4$ region.
 Our conclusion is that there is no need to calculate higher-order coefficients $d_{n\geq5}$
 if we are interested in the accuracy of the order of 1\%.
\end{abstract}

\markboth{\large \sl \underline{Alexander~P.~Bakulev} 
          \hspace*{2cm} HSQCD 2008}
         {\large \sl \hspace*{1cm} Resummation approach in FAPT: How many loops to calculate?}

\section{Basics of APT in QCD}
 \label{sec:APT}
In the standard QCD Perturbation Theory (PT) we know that 
the Renormalization Group (RG) equation $da_s[L]/dL = -a_s^2-\ldots$
for the effective coupling $\alpha_s(Q^2)=a_s[L]/\beta_f$ 
with $L=\ln(Q^2/\Lambda^2)$, $\beta_f=b_0(N_f)/(4\pi)=(11-2N_f/3)/(4\pi)$\footnote{%
   We use notations $f(Q^2)$ and $f[L]$ in order to specify what arguments we mean --- 
   squared momentum $Q^2$ or its logarithm $L=\ln(Q^2/\Lambda^2)$,
   that is $f[L]=f(\Lambda^2\cdot e^L)$ and $\Lambda^2$ is usually referred to $N_f=3$ region.}.
Then the one-loop solution generates Landau pole singularity,
$a_s[L] = 1/L$.

In the Analytic Perturbation Theory (APT) we have
different effective couplings in Minkowskian 
(Radyushkin \cite{Rad82}, and Krasnikov and Pivovarov \cite{KP82})
and Euclidean (Shirkov and Solovtsov \cite{SS}) regions.
In Euclidean domain,
$\displaystyle-q^2=Q^2$, $\displaystyle L=\ln Q^2/\Lambda^2$,
APT generates the following set of images for the effective coupling
and its $n$-th powers,  
$\displaystyle\left\{{\mathcal A}_n[L]\right\}_{n\in\mathbb{N}}$,
whereas in Minkowskian domain,
$\displaystyle q^2=s$, $\displaystyle L_s=\ln s/\Lambda^2$,
it generates another set,
$\displaystyle\left\{{\mathfrak A}_n[L_s]\right\}_{n\in\mathbb{N}}$.
APT is based on the RG and causality 
that guaranties standard perturbative UV asymptotics 
and spectral properties.
Power series $\sum_{m}d_m a_s^m[L]$ 
transforms into non-power series 
$\sum_{m}d_m {\mathcal A}_{m}[L]$ in APT.

By the analytization in APT for an observable $f(Q^2)$
we mean the ``K\"allen--Lehman'' representation
 \begin{eqnarray}
  \label{eq:An.SD}
  \left[f(Q^2)\right]_\text{an}
   = \int_0^{\infty}\!
      \frac{\rho_f(\sigma)}
         {\sigma+Q^2-i\epsilon}\,
       d\sigma
  ~~\text{with}~~
   \rho_f(\sigma)=\frac{1}{\pi}\,
                   \textbf{Im}\,
                    \big[f(-\sigma)\big]\,.
 \end{eqnarray}
Then in the one-loop approximation (note pole remover $(e^L-1)^{-1}$ in (\ref{eq:A_1}))
\begin{subequations}
 \label{eq:A.U}
 \begin{eqnarray}
  \label{eq:A_1}
 \mathcal A_1[L]
  &=& \int_0^{\infty}\!\frac{\rho(\sigma)}{\sigma+Q^2}\,d\sigma\
   =\ \frac{1}{L} - \frac{1}{e^L-1}\,,~\\
 \label{eq:U_1}  
 {\mathfrak A}_1[L_s] 
  &=& \int_s^{\infty}\!\frac{\rho(\sigma)}{\sigma}\,d\sigma\
   =\ \frac{1}{\pi}\,\arccos\frac{L_s}{\sqrt{\pi^2+L_s^2}}\,,~
 \end{eqnarray}
\end{subequations}
whereas analytic images of the higher powers ($n\geq2, n\in\mathbb{N}$) are:
\begin{eqnarray}
 \label{eq:recurrence}
 {\mathcal A_n[L] \choose \mathfrak A_n[L_s]}
  &=& \frac{1}{(n-1)!}\left( -\frac{d}{d L}\right)^{n-1}
      {\mathcal A_{1}[L] \choose \mathfrak A_{1}[L_s]}\,.
\end{eqnarray}

In the standard QCD PT we have also:\\ 
(i) the factorization procedure in QCD  
    that gives rise to the appearance of logarithmic factors of the type: 
     $a_s^\nu[L]\,L$;~\footnote{%
     First indication that a special ``analytization'' procedure
is needed to handle these logarithmic terms appeared in~\cite{KS01},
where it has been suggested that one should demand 
the analyticity of the partonic amplitude as a \textit{whole}.}\\
(ii) the RG evolution 
     that generates evolution factors of the type: 
     $B(Q^2)=\left[Z(Q^2)/Z(\mu^2)\right]$ $B(\mu^2)$, 
     which reduce in the one-loop approximation to
     $Z(Q^2) \sim a_s^\nu[L]$ with $\nu=\gamma_0/(2b_0)$ 
     being a fractional number.\\
All these means we need to construct analytic images of new functions:
$\displaystyle a_s^\nu,~a_s^\nu\,L^m, \ldots$\,.

\begin{figure}[hbt]
 \centerline{\includegraphics[width=0.45\textwidth]{
             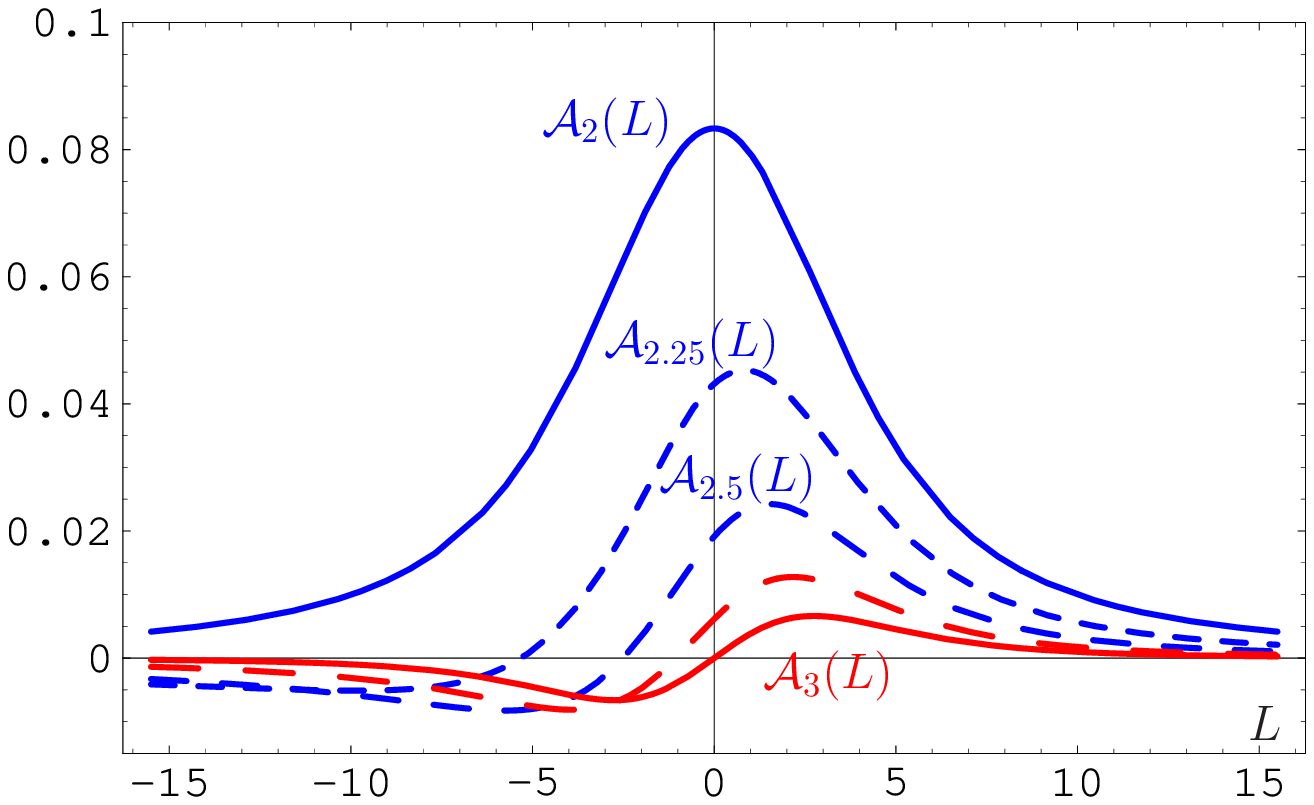}~~~%
             \includegraphics[width=0.45\textwidth]{
             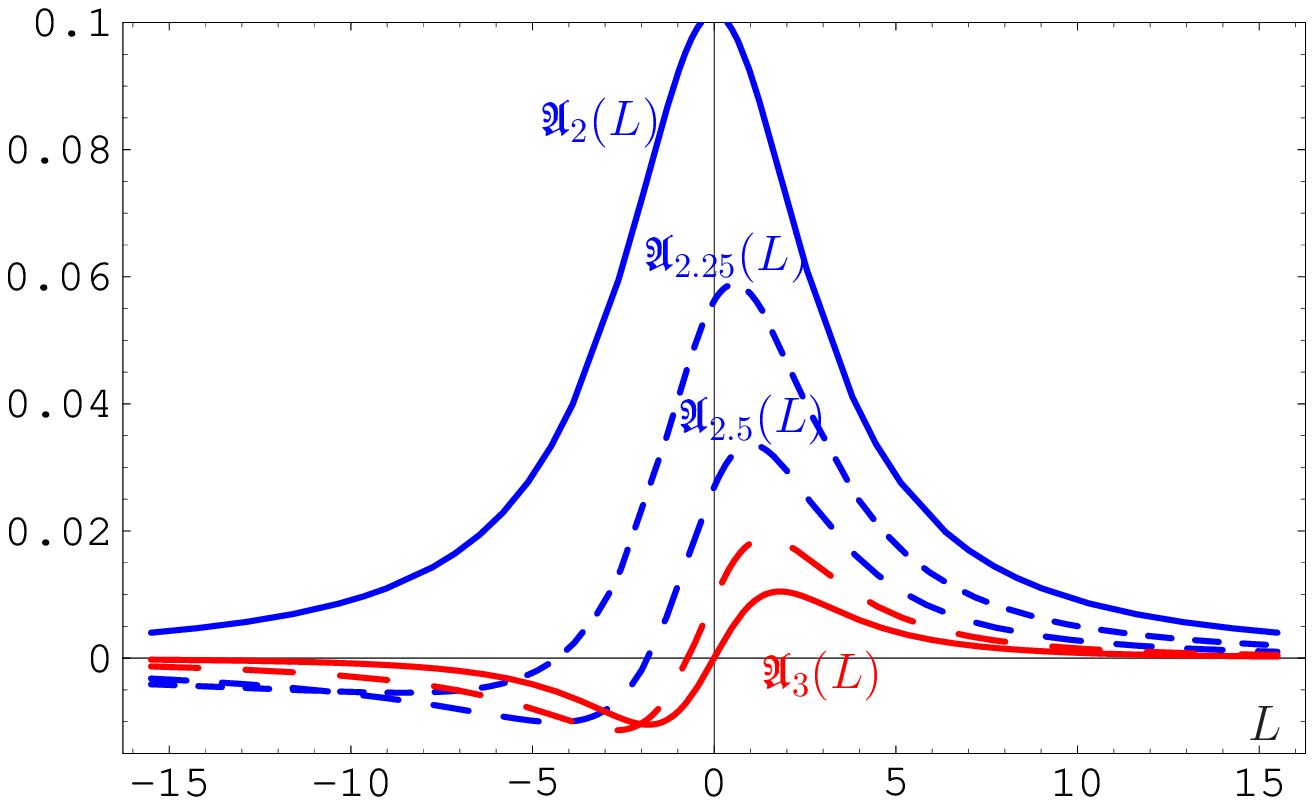}}
  \caption{Graphics of ${\mathcal A}_{\nu}[L]$ (left panel)
  and ${\mathfrak A}_{\nu}[L]$ (right panel)
  for fractional $\nu\in\left[2,3\right]$.
  \label{fig:U23_A23}}
\end{figure}

In the one-loop approximation 
using recursive relation (\ref{eq:A.U})
we can obtain explicit expressions for
${\mathcal A}_{\nu}[L]$
and ${\mathfrak A}_{\nu}[L]$:
\begin{eqnarray}
 {\mathcal A}_{\nu}[L] 
  = \frac{1}{L^\nu} 
  - \frac{F(e^{-L},1-\nu)}{\Gamma(\nu)}\,;\quad
 {\mathfrak A}_{\nu}[L] 
  = \frac{\text{sin}\left[(\nu -1)\arccos\left(L/\sqrt{\pi^2+L^2}\right)\right]}
         {\pi(\nu -1) \left(\pi^2+L^2\right)^{(\nu-1)/2}}\,.~
\end{eqnarray}
Here $F(z,\nu)$ is reduced Lerch transcendental function,
which is an analytic function in $\nu$.
Interesting to note that ${\mathcal A}_\nu[L]$ appears to be 
an entire function in $\nu$, 
whereas ${\mathfrak A}_{\nu}[L]$ 
is determined completely in terms of elementary functions.
They have very interesting properties,
which we discussed extensively in our previous papers~\cite{BMS-APT,AB08}.
Here we only display graphics of ${\mathcal A}_{\nu}[L]$ and ${\mathfrak A}_{\nu}[L]$
in Fig.\ \ref{fig:U23_A23}:
one can see here a kind of distorting mirror on both panels.

Construction of FAPT with fixed number of quark flavors, $N_f$, 
is a two-step procedure: 
we start with the perturbative result $\left[a_s(Q^2)\right]^{\nu}$,
generate the spectral density $\rho_{\nu}(\sigma)$ using Eq.\ (\ref{eq:An.SD}),
and then obtain analytic couplings 
${\mathcal A}_{\nu}[L]$ and ${\mathfrak A}_{\nu}[L]$ via Eqs.\ (\ref{eq:A.U}).
Here $N_f$ is fixed and factorized out.
We can proceed in the same manner for $N_f$-dependent quantities:
$\left[\alpha_s^{}(Q^2;N_f)\right]^{\nu}$ 
$\Rightarrow$ 
$\bar{\rho}_{\nu}(\sigma;N_f)=\bar{\rho}_{\nu}[L_\sigma;N_f]
 \equiv\rho_{\nu}(\sigma)/\beta_f^{\nu}$
$\Rightarrow$ 
$\bar{\mathcal A}_{\nu}^{}[L;N_f]$ and $\bar{\mathfrak A}_{\nu}^{}[L;N_f]$ ---
here $N_f$ is fixed, but not factorized out.
 
Global version of FAPT,
which takes into account heavy-quark thresholds,
is constructed along the same lines
but starting from global perturbative coupling
$\left[\alpha_s^{\,\text{\tiny glob}}(Q^2)\right]^{\nu}$,
being a continuous function of $Q^2$
due to choosing different values of QCD scales $\Lambda_f$,
corresponding to different values of $N_f$.
We illustrate here the case of only one heavy-quark threshold 
at $s=m_4^2$,
corresponding to the transition $N_f=3\to N_f=4$.
Then we obtain the discontinuous spectral density 
\begin{eqnarray}
 \label{eq:global_PT_Rho_4}
  \rho_n^\text{\tiny glob}(\sigma)
  =
  \rho_n^\text{\tiny glob}[L_\sigma]
   = \theta\left(L_\sigma<L_{4}\right)\,
       \bar{\rho}_n\left[L_\sigma;3\right]
    + \theta\left(L_{4}\leq L_\sigma\right)\,
       \bar{\rho}_n\left[L_\sigma+\lambda_4;4\right]\,,~~~
\end{eqnarray}
with $L_{\sigma}\equiv\ln\left(\sigma/\Lambda_3^2\right)$,
$L_{f}\equiv\ln\left(m_f^2/\Lambda_3^2\right)$
and
$\lambda_f\equiv\ln\left(\Lambda_3^2/\Lambda_f^2\right)$ for $f=4$,
which is expressed in terms of fixed-flavor spectral densities
with 3 and 4 flavors,
$\bar{\rho}_n[L;3]$ and $\bar{\rho}_n[L+\lambda_4;4]$.
However it generates the continuous Minkowskian coupling 
\begin{eqnarray}
 {\mathfrak A}_{\nu}^{\text{\tiny glob}}[L_s]
  \!&\!=\!&\! 
    \theta\left(L_s\!<\!L_4\right)
     \Bigl(\bar{{\mathfrak A}}_{\nu}^{}[L_s;3]
          -\bar{{\mathfrak A}}_{\nu}^{}[L_4;3]
          +\bar{{\mathfrak A}}_{\nu}^{}[L_4+\lambda_4;4] 
     \Bigr)
  \nonumber\\
  \!&\!+\!&\!
    \theta\left(L_4\!\leq\!L_s\right)\,
     \bar{{\mathfrak A}}_{\nu}^{}[L_s+\lambda_4;4]\,.
 \label{eq:An.U_nu.Glo.Expl}     
\end{eqnarray}
and the analytic Euclidean coupling (for more detail see in~\cite{AB08})
\begin{eqnarray}
 {\cal A}_{\nu}^{\text{\tiny glob}}[L]
  &=& \bar{{\cal A}}_{\nu}^{}[L+\lambda_4;4] 
    + \int\limits_{-\infty}^{L_4}\!
       \frac{\bar{\rho}_{\nu}^{}[L_\sigma;3]
            -\bar{\rho}_{\nu}^{}[L_\sigma+\lambda_{4};4]}
            {1+e^{L-L_\sigma}}\,
         dL_\sigma\,.
  \label{eq:Delta_f.A_nu}
\end{eqnarray}

\section{Resummation in the one-loop APT and FAPT}
\label{sec:Resum.FAPT}
We consider now the perturbative expansion
of a typical physical quantity,
like the Adler function and the ratio $R$, 
in the one-loop APT 
\begin{eqnarray}
 \label{eq:APT.Series}
  {\mathcal D[L]\choose \mathcal R[L]}
   = d_0
   + \sum_{n=1}^{\infty}
      d_n\,{\mathcal A_{n}[L]\choose \mathfrak A_{n}[L]}\,.
\end{eqnarray}
We suggest that there exist the generating function $P(t)$
for coefficients $\tilde{d}_n=d_n/d_1$:
\begin{equation}
 \tilde{d}_n
  =\int_{0}^\infty\!\!P(t)\,t^{n-1}dt
   ~~~\text{with}~~~
   \int_{0}^\infty\!\!P(t)\,d t = 1\,.
 \label{eq:generator}
\end{equation}
To shorten our formulae, we use the following notation
$\langle\langle{f(t)}\rangle\rangle_{P(t)}\equiv\int_{0}^{\infty}\!\!f(t)P(t)dt$.
Then coefficients $d_n = d_1\,\langle\langle{t^{n-1}}{P(t)}$
and as has been shown in~\cite{MS04}
we have the exact result for the sum in (\ref{eq:APT.Series})  
\begin{eqnarray}
 \label{eq:APT.Sum.DR[L]}
  {\mathcal D[L]\choose \mathcal R[L]}
   = d_0 + d_1\,{\langle\langle{\mathcal A_1[L-t]}\rangle\rangle_{P(t)}
                  \choose
                   \langle\langle{\mathfrak A_1[L-t]}\rangle\rangle_{P(t)}}\,.
\end{eqnarray}
The integral in variable $t$ here has a rigorous meaning, 
ensured by the finiteness of the couplings $\mathcal A_1[t] \leq 1$
and  $\mathfrak A_1[t] \leq 1$
and fast fall-off of the generating function $P(t)$.

In our previous publications~\cite{AB08,BM08}
we have constructed generalizations of these results,
first, to the case of the global APT,
when heavy-quark thresholds are taken into account.
Then one starts with the series 
of the type (\ref{eq:APT.Series}),
where $\mathcal A_{n}[L]$ or $\mathfrak A_{n}[L]$
are substituted by their global analogs,
$\mathcal A_{n}^\text{\tiny glob}[L]$ or
$\mathfrak A_{n}^\text{\tiny glob}[L]$
(note that due to different normalizations of global
 couplings, $\mathcal A_{n}^\text{\tiny glob}[L]\simeq\mathcal A_{n}[L]/\beta_f$,
 the coefficients $d_n$ should be also changed).
The most simple generalization of the summation result
appears in Minkowski domain:
\begin{eqnarray}
 \mathcal R^\text{\tiny glob}[L]
  =  d_0
 \!\!&\!+\!&\!\! d_1 \langle\langle{
          \theta(L\!<\!L_4)
           \left[\Delta_{4}\bar{\mathfrak A}_{1}[t]
                +\bar{\mathfrak A}_{1}\!\Big[L\!-\!\frac{t}{\beta_3};3\Big]
           \right]     }\rangle\rangle_{P(t)}\nonumber\\
 \!\!&\!+\!&\!\! d_1 \langle\langle{
          \theta(L\!\geq\!L_4)
           \bar{\mathfrak A}_{1}\!\Big[L\!+\!\lambda_4-\!\frac{t}{\beta_4};4\Big]
                       }\rangle\rangle_{P(t)}\,;~~~
 \label{eq:sum.R.Glo.4}
\end{eqnarray}
where $\Delta_4\bar{\mathfrak A}_\nu[t]\equiv
  \bar{\mathfrak A}_\nu\!\Big[L_4+\lambda_{4}-t/\beta_4;4\Big]
 -\bar{\mathfrak A}_\nu\!\Big[L_3-t/\beta_3;3\Big]$.
 
The second generalization has been obtained for the case 
of the global FAPT.
Then the starting point is the series of the type 
$\sum_{n=0}^{\infty} d_n\,\mathfrak A_{n+\nu}^\text{\tiny glob}[L]$
and the result of summation is a complete analog of Eq.\ (\ref{eq:sum.R.Glo.4})
with substitutions
\begin{eqnarray}
 \label{eq:P_nu(t)}
  P(t)\Rightarrow P_{\nu}(t) = 
   \int_0^{1}\!P\left(\frac{t}{1-x}\right)
    \frac{\nu\,x^{\nu-1}dx}
         {1-x}\,,
\end{eqnarray}         
$d_0\Rightarrow d_0\,\bar{\mathfrak A}_{\nu}[L]$, 
$\bar{\mathfrak A}_{1}[L-t]\Rightarrow 
 \bar{\mathfrak A}_{1+\nu}[L-t]$, 
and
$\Delta_4\bar{\mathfrak A}_{1}[t]\Rightarrow 
 \Delta_4\bar{\mathfrak A}_{1+\nu}[t]$.
Needless to say that all needed formulas have been also obtained 
in parallel for the Euclidean case.

\section{Applications to Higgs boson decay and Adler function}
\label{sec:Appl.Higgs}
First, we analyze the Higgs boson decay to a $\bar{b}b$ pair.
Here we have for the decay width  
\begin{eqnarray}
 \Gamma(\text{H} \to b\bar{b})
  = \frac{G_F}{4\sqrt{2}\pi}\,
     M_{H}\,
      \widetilde{R}_\text{\tiny S}(M_{H}^2)
  ~~~\text{with}~~~
  \widetilde{R}_\text{\tiny S}(M_{H}^2)
  \equiv m^2_{b}(M_{H}^2)\,R_\text{\tiny S}(M_{H}^2)
 \label{eq:Higgs.decay.rate}
\end{eqnarray}
and
$R_\text{\tiny S}(s)$ 
is the $R$-ratio for the scalar correlator,
see for details in~\cite{BMS-APT,BCK05}.
In the one-loop FAPT this generates the following
non-power expansion\footnote{%
Appearance of denominators $\pi^n$ in association
with the coefficients $\tilde{d}_n$
is due to $d_n$ normalization.}:
\begin{eqnarray}
 \widetilde{\mathcal R}_\text{\tiny S}[L]
   =  3\,\hat{m}_{(1)}^2\,
      \Bigg\{\mathfrak A_{\nu_{0}}^{\text{\tiny glob}}[L]
          + d_1^\text{\,\tiny S}\,\sum_{n\geq1}
             \frac{\tilde{d}_{n}^\text{\,\tiny S}}{\pi^{n}}\,
              \mathfrak A_{n+\nu_{0}}^{\text{\tiny glob}}[L]
      \Bigg\}\,,
 \label{eq:R_S-MFAPT}
\end{eqnarray}
where $\hat{m}_{(1)}^2$ is the renormalization-group
invariant of the one-loop $m^2_{b}(\mu^2)$ evolution
$m_{b}^2(Q^2) = \hat{m}_{(1)}^2\,\alpha_{s}^{\nu_{0}}(Q^2)$
with $\nu_{0}=2\gamma_0/b_0(5)=1.04$ and
$\gamma_0$ is the quark-mass anomalous dimension
(for a discussion --- see in~\cite{KK08}).

\begin{figure}[hb]
 \centerline{\includegraphics[width=0.45\textwidth]{
             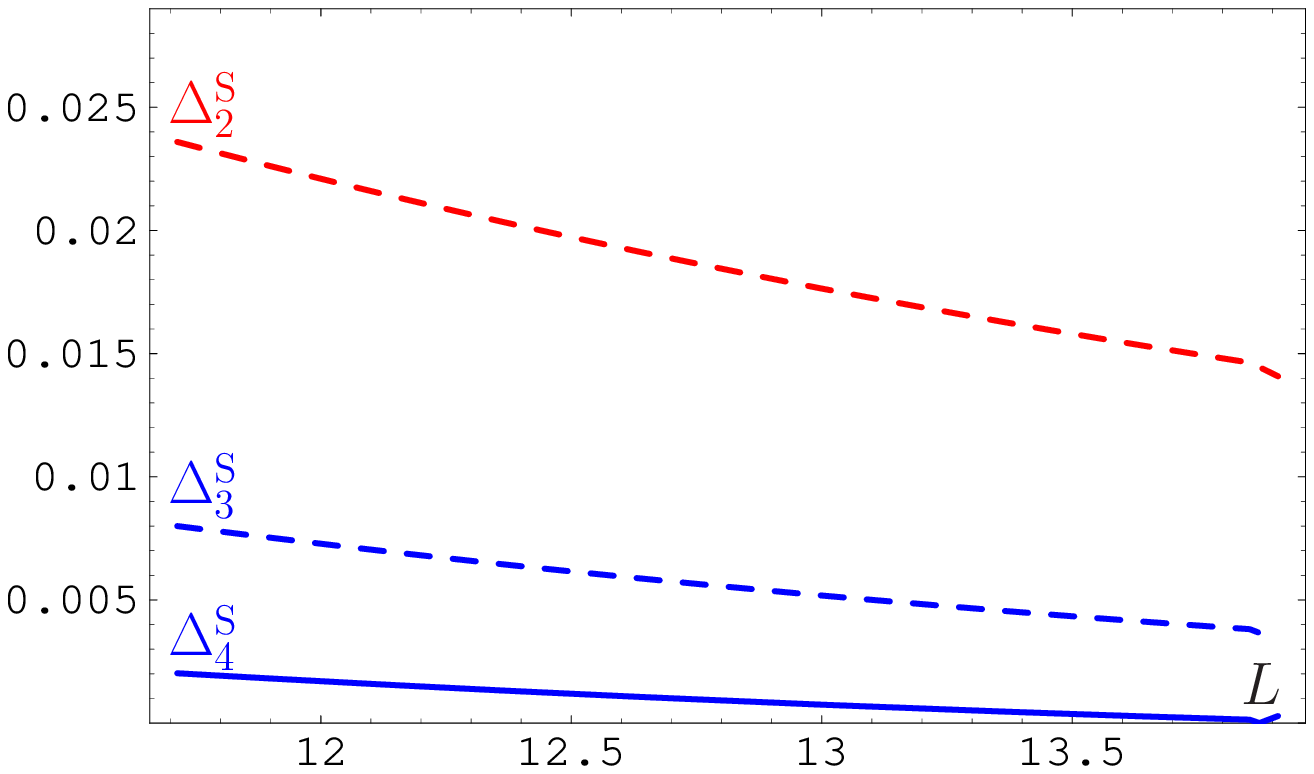}
          ~~~\includegraphics[width=0.45\textwidth]{
             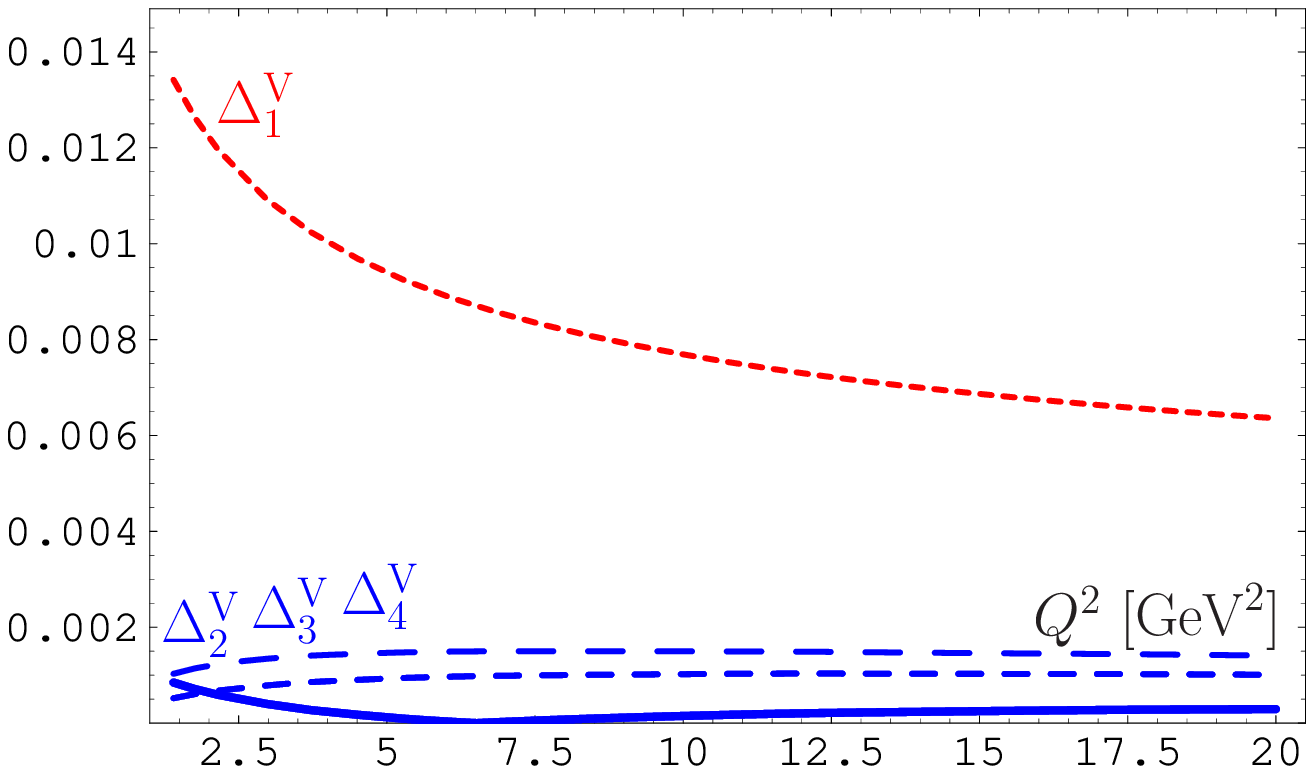}
  \vspace*{-1mm}}
   \caption{\textbf{Left panel:} The relative errors  $\Delta^\text{S}_N[L]$, $N=2, 3$
    and $4$, of the truncated FAPT, Eq.\ (\ref{eq:FAPT.trunc}), 
    in comparison with the exact summation result, Eq.\ (\ref{eq:R_S.Sum}).
    \textbf{Right panel:} Analogous relative errors $\Delta^\text{V}_N(Q^2)$, 
    $N=1, \ldots, 4$, for the case of vector Adler function
    (solid line is for $N=2$, dashed --- for $N=3$, and long-dashed --- for $N=4$).
   \label{fig:Higgs}}
\end{figure}

We take for the generating function $P(t)$
the Lipatov-like model of~\cite{BM08,Lip76} 
with $\left\{c=2.4,~\beta=-0.52\right\}$
\begin{eqnarray}
\label{eq:Higgs.Model}
  \tilde{d}_{n}^\text{\,\tiny S}
   = c^{n-1}\frac{\Gamma (n+1)+\beta\,\Gamma (n)}{1+\beta}\,;\quad
  P_\text{\tiny S}(t)
  = \frac{(t/c)+\beta}{c\,(1+\beta)}\,e^{-{t/c}}\,.
\end{eqnarray}
It gives a very good prediction for
$\tilde{d}_{n}^\text{\,\tiny S}$ with $n=2, 3, 4$,
calculated in the QCD PT~\cite{BCK05}:
$7.50$, $61.1$, and  $625$
in comparison with
$7.42$, $62.3$, and  $620$.
Then we apply FAPT resummation technique
to estimate
how good is FAPT
in approximating the whole sum $\widetilde{\mathcal R}_\text{\tiny S}[L]$
in the range $L\in[11,13.8]$
which corresponds to the range
$M_H\in[60,170]$~GeV$^2$
with $\Lambda^{N_f=3}_{\text{QCD}}=172$~MeV
and ${\mathfrak A}^{\text{\tiny glob}}_{1}(m_Z^2)=0.120$.
In this range we have ($L_6=\ln(m_t^2/\Lambda_3^2)$)
\begin{eqnarray}
 \widetilde{\mathcal R}_\text{\tiny S}[L]
   = 3\,\hat{m}_{(1)}^2\,
       \left\{{\mathfrak A}^\text{\tiny glob}_{\nu_{0}}[L]
           + \frac{d_{1}^\text{\,\tiny S}}{\pi}\,
              \langle\langle{\bar{\mathfrak A}_{1+\nu_{0}}\!
                \Big[L\!+\!\lambda_5\!-\!\frac{t}{\pi\beta_5};5\Big]
                \!+\!
                \Delta_{6}\bar{\mathfrak A}_{1+\nu_{0}}
                 \left[\frac{t}{\pi}\right]
                            }\rangle\rangle_{P_{\nu_{0}}^\text{\,\tiny S}}
       \right\}
 \label{eq:R_S.Sum}
\end{eqnarray}
with $P_{\nu_{0}}^\text{\,\tiny S}(t)$ defined via Eqs.\ (\ref{eq:Higgs.Model})
and (\ref{eq:P_nu(t)}).
Now we analyze the accuracy of the truncated FAPT expressions
\begin{eqnarray}
 \label{eq:FAPT.trunc}
 \widetilde{\mathcal R}_\text{\tiny S}[L;N]
  &=& 3\,\hat{m}_{(1)}^2\,
       \left[{\mathfrak A}_{\nu_{0}}^{\text{\tiny glob}}[L]
           + d_1^\text{\,\tiny S}\,\sum_{n=1}^{N}
              \frac{\tilde{d}_{n}^\text{\,\tiny S}}{\pi^{n}}\,
               {\mathfrak A}_{n+\nu_{0}}^{\text{\tiny glob}}[L]
       \right]
\end{eqnarray}
and compare them with the total sum
$\widetilde{\mathcal R}_\text{\tiny S}[L]$
in Eq.\ (\ref{eq:R_S.Sum})
using relative errors
$\Delta_N^\text{S}[L]=1-\widetilde{\mathcal R}_\text{\tiny S}[L;N]/\widetilde{\mathcal R}_\text{\tiny S}[L]$.
In the left panel of Fig.\ \ref{fig:Higgs}
we show these errors for $N=2$, $N=3$, and $N=4$
in the analyzed range of $L\in[11,13.8]$.
We see that already $\widetilde{\mathcal R}_\text{\tiny S}[L;2]$
gives accuracy of the order of 2.5\%,
whereas $\widetilde{\mathcal R}_\text{\tiny S}[L;3]$
of the order of 1\%.
That means that there is no need to calculate further corrections:
at the level of accuracy of 1\% it is quite enough to take into account
only coefficients up to $d_3$.
This conclusion is stable
with respect to the variation of parameters
of the model $P_\text{\tiny S}(t)$.

Next, we analyze the Adler function of vector correlator.
In the one-loop APT this generates the following
non-power expansion:
\begin{eqnarray}
 \mathcal D_\text{\tiny V}[L]
   = 1 + d_1^\text{\tiny V}\,\sum_{n\geq1}
             \frac{\tilde{d}_{n}^\text{\tiny V}}{\pi^{n}}\,
              \mathcal A_{n}^{\text{\tiny glob}}[L]\,,
 \label{eq:D_V-APT}
\end{eqnarray}
Here we use another Lipatov-like model for perturbative coefficients 
in the $N_f=4$ region
\begin{eqnarray}
\label{eq:Adler.Model}
  \tilde{d}_n^\text{\tiny V}
   = c^{n-1}\,
     \frac{\beta^{n+1}-n}
          {\beta^2-1}\,\Gamma(n)\,;\quad
  P_\text{\tiny V}(t)
  = \frac{\beta\,e^{-t/c\beta} - (t/c)\,e^{-t/c}}
           {c\left(\beta ^2-1\right)}
\end{eqnarray}
with $c=3.456$ and $\beta=1.325$,
which gives a very good prediction for
$\tilde{d}_n^\text{\tiny V}$ with $n=2, 3, 4$,
calculated in the QCD PT~\cite{BCK08}:
$1.49$, $2.60$, and  $27.5$
in comparison with
$1.52$, $2.59$, and  $27.4$.
To estimate 
how good is APT
in approximating the whole sum $\mathcal D_\text{\tiny S}[L]$,
we apply APT resummation approach 
in the range $Q^2\in[1.5,20]$~GeV$^2$,
corresponding to the $N_f=4$ value.
Again we analyze the accuracy of the truncated APT expressions
$\mathcal D_\text{\tiny V}[L;N]= 1 + d_1^\text{\,\tiny V}\,\sum_{n=1}^{N}
           \frac{\tilde{d}_{n}^\text{\,\tiny V}}{\pi^{n}}\,
            \mathcal A_{n}^{\text{\tiny glob}}[L]$
and compare them with the total sum
$\mathcal D_\text{\tiny V}[L]$,
obtained by resummation APT method,
using relative errors
$\Delta_N^\text{V}[L]=1-\mathcal D_\text{\tiny V}[L;N]/\mathcal D_\text{\tiny V}[L]$.
In the right panel of Fig.\ \ref{fig:Higgs}
we show these errors for $N=1,\ldots, 4$
in the analyzed range of $Q^2$.
We see that already $\mathcal D_\text{\tiny V}(Q^2;2)$
gives the accuracy of the order of 0.05\%,
whereas taking into account higher-order corrections 
only worsen the accuracy: 
$\mathcal D_\text{\tiny V}(Q^2;3)$ provides
the accuracy of the order of 0.1\%
and $\mathcal D_\text{\tiny V}(Q^2;4)$
--- of the order of 0.2\%.
That means that the NLO approximation gives 
the best result and after that 
the series starts to reveal its asymptotic character.

\section{Conclusions}
\label{sec:Concl}
In this report we described the resummation approach 
in the global versions of the one-loop APT and FAPT
and argued
that it produces finite answers in both Euclidean and Minkowski regions, 
provided the generating function $P(t)$ 
of perturbative coefficients $d_n$ is known.
In the case of the Higgs boson decay 
an accuracy of the order of 1\% is reached at N$^3$LO approximation, 
when term $d_3{\mathcal A}_3$ is taken into account,
whereas for the Adler function $D(Q^2)$ 
we have an accuracy of the order of $0.1$\% already at N$^2$LO
(i.e., with taking into account $d_2{\mathcal A}_2$ term). 

The main conclusion is: 
In order to achieve an accuracy of the order of 1\% 
we do not need to calculate more than four loops and 
$d_4$ coefficients are needed only to estimate 
corresponding generating functions $P(t)$.

\section*{Acknowledgements} 
I would like to thank the organizers of the Conference ``Hadron Structure and QCD--08'' 
(Gatchina, Russia, June 30--July 4, 2008) for the invitation and support.
This work was supported in part by 
the Russian Foundation for Fundamental Research, 
grants No.\ 06-02-16215, 07-02-91557, and 08-01-00686,
the BRFBR--JINR Cooperation Programme (contract No.\ F06D-002),
the Heisenberg--Landau Programme under grant 2008, 
and the Deutsche Forschungsgemeinschaft
(project DFG 436 RUS 113/881/0).


\end{document}